We consider the optical conductivity of carriers scattered by local sombrero potentials associated with hindered planar rotators in solids. Examples for hindered rotators are provided by off-center substitutional $Li^+$ ions in colored alkali halides and off-site apical oxygens O(A) in layered perovskites. The off-site displacements are obtained by solving for the vibronic mixing of electronic states at the displaced ion. It turns out that while linear mixing terms produce off-site displacements, cubic terms produce reorientational barriers to hinder the free rotation of off-center ions around their normal lattice sites. The eigenvalue equation of a hindered rotator is solved in 2-D by Mathieu's periodic functions, while the rotation is quantized in rotational bands. We discuss the optical conductivities pertinent to each of the exemplified cases.

1. Introduction

Under certain conditions, impurities or even host ions in solids may go slightly off-site relative to their normal lattice positions. Examples are provided by a variety of cases ranging from substitutional $Li^+$ ions in alkali halides to apical oxygen ions O(A) in layered perovskites [1-5]. Such ions equilibrate off-center, due to site-splitting potentials whose origin has been ascribed to polarization-mediated attraction and orbital-overlap repulsion between the impurity and surrounding host ions balancing each other off-site [6]. At the same time, the quantum-mechanical approach has attributed the site-splitting potentials to the pseudo-Jahn-Teller effect (PJT) [7]. The latter effect is a local lattice instability, arising from the strong mixing of nearly-degenerate electronic states of the impurity-containing cluster by an appropriate vibrational mode. The off-site ion may perform reorientational hops around the central site which resemble a hindered rotation amongst reorientational wells separated by interwell barriers. While a site-splitting potential arises as a first-order effect in the electron-vibrational mode coupling, the interwell multibarrier potential is third order in that coupling [8,9]. In most of the known cases the hindered rotator is three-dimensional (3-D), such as the off-center rotation of isolated $Li^+$ impurities in KCl [10]. There are a few examples of two-dimensional rotation (2-D), such as the one performed by off-center $Li^+$ ions nearest-neighboring F centers to give $F_A(Li)$ centers in colored alkali halides [10]. A planar off-site rotation might also be exhibited by O(A) in layered perovskites if these ions coupled strongly to $E_u$ phonon modes in tetragonal symmetry [11]. One-dimensional (1-D) flip-flops along the c-axis are conceivable for O(A) via the coupling to the $A_{2u}$ mode in $La_{2-x}Sr_xCuO_4$ [12].

It may be expected that the conventional *spherical* form of a rotating off-center ion will only be preserved in high-symmetry cubic lattices, such as the alkali halides. For lower symmetry lattices, such as tetragonal or orthorhombic layered perovskites, the shape of a rotating ion may change to *oblate*, if the rotational axis is perpendicular to the ion's symmetry axis, or to *prolate*, if the rotational axis is parallel to the ion's symmetry axis. However, even in cubic symmetry there may be inter-conversions from one form to the other, possibly accompanied by respective changes of rotational barriers as the temperature is raised [2]. It may be worth noting that in the respective inter-conversion of shape from oblate to prolate in nuclear physics the rotational band structure is disrupted by the transition from a collective rotational state to a single-particle state [13].

Another example of the oblate (spheroid) shape persisting in an otherwise isotropic space is rotating Earth's where the asymmetry arises due to the centrifugal force which grows stronger as one moves towards the equator. It is plausible that the smeared ion density of an off-center rotator at rotational sites along the Mexican Hat brim may undergo similar centrifugal pressure to turn oblate, even though in a high-symmetry cubic lattice.

A variety of physical properties are affected by impurity or host ions going off-site and performing hindered reorientational jumps. Among these are the spin-lattice relaxation rate and the charge-carrier scattering. A key quantity controlling both of them is the relaxation rate along the reorientational path or trajectory [12]. Another off-center dependent property is the optical absorption or conductivity spectrum. While the spin-lattice relaxation and electric transport controlled by site-splitting potentials have been addressed elsewhere [14,15], the present paper will focus on the optical spectra. Preliminary insights into the problem have been reported in the literature, even though relating to 1D flip-flops mainly dealt with by means of approximate harmonic-oscillator solutions to the multi-dimensional problem [16,17].

Hopefully, optical conductivity spectra of hindered rotators may provide optical signatures to be used for identifying the off-center species. This may prove valuable if off-center displacements have not yet been proven unambiguously, such as ones in the tetragonal or orthorhombic $La_{2-x}Sr_xCuO_4$ family. We stress that our principle conclusions on the off-center ions will eventually apply to the vibronic small polarons as well, in so far as both ion and carrier species produce similar distortion patterns.

## 2. Hamiltonian

Site-splitting potentials associated with off-center impurity ions in alkali halides or apex oxygens in layered perovskites have been evidenced experimentally, even though there is no commonly accepted convention on their quantum-mechanical origin [18]. Throughout this paper, we assume that the site-splitting potentials originate from the pseudo-Jahn-Teller (PJT) mixing of nearly-degenerate different-parity electronic states by an odd parity phonon mode of a cluster containing the off-center species [8,9]. To be specific, we consider $|a_{1g}> \equiv |s>$ and $|t_{1ui}> \equiv |p_i>$ electronic states coupled to a $T_{1u}$ phonon mode in $O_h$ symmetry, as described by the following local PJT Hamiltonian:

$$H = \sum_{n\alpha}\varepsilon_{n\alpha} a_{n\alpha}^\dagger a_{n\alpha} + \sum_n \sum_{\alpha\neq\beta}\{\sum_i G_i Q_i + \sum_{ijk} B_{ijk} Q_i Q_j Q_k\}_n a_{n\alpha}^\dagger a_{n\beta} +$$

$$\tfrac{1}{2}\sum_n (\boldsymbol{P}_n^2/M_n + K_n Q_n^2) \qquad (1)$$

where n is the site label, i,j,k = x,y,z, $\alpha,\beta = a_{1g}, t_{1ui}$ are electron energy level labels, $a_{n\alpha}^\dagger$ ($a_{n\alpha}$) are electron creation (annihilation) operators in the site representation, $\varepsilon_{n\alpha}$ are electron energy terms. $G_i$ and $B_{ijk}$ are the electron-mode coupling constants, first- and third-order, respectively, $\boldsymbol{P}_n$, $K_n = M_n \omega_{bare}^2$, $M_n$, and $\omega_{bare}$ are correspondingly the momentum, spring constant, reduced mass, and bare vibrational frequency of the coupled phonon mode with coordinates $Q_{in}$, etc., the latter regarded as c-numbers. Locality implies that hopping terms of the form $\sum_{n\alpha} t_{nn\pm1\alpha} (a_{n+1\alpha}^\dagger - a_{n-1\alpha}^\dagger) a_{n\alpha}$ are presently dropped from consideration.

Setting $\varepsilon_{n\alpha} = -\varepsilon_{n\beta i} = \tfrac{1}{2}\varepsilon_{n\alpha\beta}$, we get the following adiabatic eigenvalues, site labels dropped, as uniformity over the sites being assumed for simplicity:

$$\varepsilon_{U/L}(Q) = \tfrac{1}{2}\sum_i K_i Q_i^2 \pm \tfrac{1}{2}\sqrt{[4\sum_i (G_i Q_i + \sum_{jk} B_{ijk} Q_i Q_j Q_k)^2 + \varepsilon_{\alpha\beta}^2]} \qquad (2)$$

where the signs upper (+) and lower (−) refer to the upper (U) and lower (L) energy branches, respectively. Isotropic behavior is assumed along the crystallographic axes in cubic symmetry, while the $B_{ijk}$ tensor is simplified leading to [19]:

$$\varepsilon_{U/L}(Q) \approx \tfrac{1}{2}KQ^2 \pm \sqrt{\{(GQ)^2 + 2G[(B_c - B_b)\sum_i Q_i^4 + B_b Q^4] + (\varepsilon_{\alpha\beta}/2)^2\}}, \qquad (3)$$

$$\varepsilon_{U/L}(Q_0) \approx \tfrac{1}{2}KQ_0^2 \pm \sqrt{\{(GQ_0)^2 + (\varepsilon_{\alpha\beta}/2)^2\}} \pm G[(B_c - B_b)\sum_i Q_I^4 + B_b Q^4] /$$

$$\sqrt{\{(GQ_0)^2 + (\varepsilon_{\alpha\beta}/2)^2\}} =$$

$$\pm (K/G)[(B_c - B_b)\sum_i Q_I^4 + B_b Q^4] + \varepsilon_{JT}[(1\pm 2) - (\varepsilon_{\alpha\beta}/4\varepsilon_{JT})^2] \qquad (4)$$

at $Q \approx Q_0 = (4G^4 - K^2 \varepsilon_{\alpha\beta}^2)^{1/2} / 2GK$, $\varepsilon_{JT} = G^2/2K$. Here $Q = \sum_i Q_i^2$ *is* the mode radial coordinate, while the simplified B-tensor is $B_b = B_{ijj}$, $B_c = B_{iii}$, $B_{ijk} = 0$ at $i\neq j\neq k\neq i$. The adiabatic eigenvalues (3) in 2-D are informative at all B = 0 (first-order coupling) or at some B ≠ 0 (coupling up to third-order).

To first order in the electron-mode coupling at all $B_{ijk} \equiv 0$, the lower branch is double well at $\pm Q_0 = \pm\sqrt{(2\varepsilon_{JT}/K)}\sqrt{(1-\eta^2)}$, with $\eta = \varepsilon_{\alpha\beta}/4\varepsilon_{JT}$, for $\eta<1$ *i.e.* $4\varepsilon_{JT} \equiv 2G^2/K > \varepsilon_{\alpha\beta}$. We note that the interwell transfer band lying at $\varepsilon = -\varepsilon_b = -\varepsilon_{JT}(1+\eta^2) + \tfrac{1}{2}\hbar\omega$, ($\varepsilon_{JT}$ - Jahn-Teller's energy), $\varepsilon_b$ enters as an off-center small polaron binding energy so that $\eta$ is a measure of the polaron size: Polarons with $\eta \ll 1$ are small, those with $\eta \leq 1$ are large, the remaining ones are of the intermediate size. The upper branch is a single anharmonic well centered at $Q = 0$.

## 3. Cluster calculations

In order to reveal the electronic energy levels most likely involved in the vibronic mixing interaction, we carried out cluster calculations by the Extended Hückel

Method [20]. We first reproduced the details of an earlier calculation of the energies of a cluster composed of three Cl$^-$ pairs nearest-neighboring a Li$^+$ ion in KCl and found them qualitatively in concert [9]. In particular, we found a pair of two opposite-parity states of symmetry $a_{1g}$ and $t_{1u}$, respectively, that split by $\approx 7$ eV when Li$^+$ was on-center. However, this gap was nearly twice as large as the one reported earlier. Subjecting Li$^+$ to small displacements along <111> we obtained minima and maxima in $t_{1u}$ and $a_{1g}$, respectively, which reproduced to a reasonable extent at small $Q$ the square-root parts of the adiabatic energies $\varepsilon_{U/L}(Q)$. Fitting to the theoretical equation for $\varepsilon_{U/L}(Q)$, we roughly estimated $G \approx 20$ eV/Å at $\varepsilon_{\alpha\beta} = 7$ eV.

We also considered a cube composed of eight La$^{3+}$ ions enclosing a Cu$^{2+}$.6O$^{2-}$ octahedron in La$_{2-x}$Sr$_x$CuO$_4$ assuming the ionic valent charges. However, the total cluster charge could be varied depending on the goals of the particular calculation. Thus, a total charge of $q = +14$ was the sum of the ionic charges in an undisturbed cluster, while $q = +15$ was the total charge of a cluster to which an extra hole was added. Varying the total charge did not change the energy level spectrum but only the level populations.

For $q = +14$, there is an uppermost unpaired 3b$_{1g}$ electron at $\varepsilon_{b1} = -11.524$ eV, while the subsequent higher-lying 4a$_{2u}$ level at $\varepsilon_{a2} = -9.099$ eV is empty. The respective energy-level separation of $\Delta_{b1-a2} = 2.425$ eV may be favorable for a PJT mixing by the A$_{2u}$ mode: At $K = 1.624$ eV/Å$^2$ of a two-atom oxygen vibrator at $\Omega = 235$ cm$^{-1}$, a strong PJT mixing would occur for $G > 1.403$ eV/Å. For $q = +15$, the extra hole would normally reside on the 3b$_{1g}$ orbital. Now, the same PJT mixing will persist if one of the two lower-lying 5a$_{1g}$ electrons at $\varepsilon_{a1} = -13.317$ eV is lifted to the 3b$_{1g}$ state at $\Delta_{a1-b1} = 1.793$ eV above, that is, if the extra hole resides on the 5a$_{1g}$ orbital. Indeed, this is the prerequisite for the interlayer hole leak. The 8La$^{3+}$.Cu$^{2+}$.6O$^{2-}$ cluster being a basic building block, our numerical calculations provide rough estimates for the b$_1$ and a$_1$ energy gaps of the La$_2$CuO$_4$ crystal.

### 4. Hindered rotators

Although hindered rotators in solids have been considered in the literature for long, some progress has been made in elucidating their quantum-mechanical behavior [21]. We illustrate the method by considering a specific system, $|a_{1g}>$ and $|t_{1ui}>$ electronic states coupled to a $T_{1u}$ phonon mode in cubic $O_h$ site-symmetry, though the results obtained herein will be extendable to other symmetries. At this stage, no coupling of the $t_{1ui}$ triply-degenerate levels to the $T_{2g}$ phonon mode is introduced, as done elsewhere.

The parametric isotropic characteristics of cubic symmetry ($K_i = K$, $G_i = G$, $\varepsilon_{\alpha\beta i} = \varepsilon_{\alpha\beta}$) will be assumed to preserve the symmetric spherical shape of the rotating ion. The oblate and prolate shapes obtain e.g. for $G_x = G_y \ll G_z$ and $G_z \ll G_x = G_y$, respectively, while all the other parameters remain isotropic. In other words, the shape is solely dependent on the rotating particle's coupling to the boson field (through boson mixing of fermion states, to be exact).

Inserting $\varepsilon_{U/L}(Q)$ into the vibronic energy operator at $Q = Q_0$ and introducing spherical coordinates at constant radius $Q_0$, we get the Hamiltonian of a *3-D hindered rotator*:

$$H_{vib}{}^{3D} = -(\hbar^2/2I)\Delta_{\theta,\varphi} + \varepsilon_{U/L}(Q)$$

$$= -(\hbar^2/2I)\Delta_{\theta,\varphi} \pm (I\omega^2/G)Q_0^2\{(B_c-B_b)[(\cos\varphi\sin\theta)^4+(\sin\varphi\sin\theta)^4+(\cos\theta)^4] + B_b\}$$

$$+ \varepsilon_{JT}[(1\pm2) - (\varepsilon_{\alpha\beta}/4\varepsilon_{JT})^2] \quad (5)$$

Discarding the radial derivative on going from the general Hamiltonian (1) to the one of a rigid rotator (5) implies that the vibronic potential is extremal and slowly varying near $Q \approx Q_0$. Hereafter $I = MQ_0^2$ is the inertial moment of the rotating species. The eigenstates and eigenvalues of the *3-D* hindered rotator are unknown. Four out of a total of eight reorientational sites in *3-D* remain active in *2-D*, as one of the mode coordinates ($Q_z$) is frozen. Most of the subsequent discussion will focus onto the spherical case.

The eigenvalue equation of a *2-D* rotator obtains by setting $\theta = \frac{1}{2}\pi$ in $H_{vib}{}^{3D}$:

$$H_{vib}{}^{2D} = -(\hbar^2/2I)(\partial^2/\partial\varphi^2) \pm (I\omega^2/G)Q_0^2\{(B_c - B_b)[(\cos\varphi)^4 + (\sin\varphi)^4] + B_b\}$$

$$+ \varepsilon_{JT}[(1\pm2) - (\varepsilon_{\alpha\beta}/4\varepsilon_{JT})^2]$$

$$= -(\hbar^2/2I)(\partial^2/\partial\varphi^2) \pm (I\omega^2/G)Q_0^2[(B_c - B_b)(3 + \cos4\varphi)/4 + B_b]$$

$$+ \varepsilon_{JT}[(1\pm2) - (\varepsilon_{\alpha\beta}/4\varepsilon_{JT})^2] \quad (6)$$

The *2-D* equation is solved in transcendent periodic functions, while its eigenvalues quantize in rotational energy bands [9]. Introducing the vibrational frequency renormalized by the electron-mode coupling

$$\omega_{ren} = \omega_{bare}[4(B_b-B_c)/G]^{1/2}Q_0 = \omega_{bare}[1-(\varepsilon_{\alpha\beta}/4\varepsilon_{JT})^2]^{1/2}[8\varepsilon_{JT}(B_b-B_c)/GK]^{1/2}, B_b > B_c \quad (7)$$

where $\omega_{bare} = \omega$ is the bare vibrational frequency, and the reorientational barrier

$$\varepsilon_{rot} = I\omega_{ren}^2/8, \quad (8)$$

the eigenvalue equation of a *2-D* hindered rotator reads:

$$-(\hbar^2/2I)(d^2\chi/d\varphi^2) + 2B_\pm\cos(4\varphi)\chi + (C_\pm - E)\chi = 0$$

$$-B_\pm = \pm\varepsilon_{rot}/4$$

$$-C_\pm = \pm\tfrac{1}{2}[(3B_c+B_b)/(B_b-B_c)]\varepsilon_{rot} - \varepsilon_{JT}[(1\pm2) - (\varepsilon_{\alpha\beta}/4\varepsilon_{JT})^2]. \quad (9)$$

The rotational eigenstates are:

$$\chi(\varphi,q) = Y(z,q)$$

where $Y(z,q)$ are solutions to Mathieu's equation

$$d^2Y/dz^2 + (a - 2q\cos(2z))Y = 0$$

$$q = 2B_{\pm}I/\hbar^2 = \pm(2\varepsilon_{rot}/\hbar\omega_{ren})^2$$

$$z = 2\varphi \tag{10}$$

where $Y(z,q)$ are Mathieu's periodic functions. The eigenvalues are two types:

$$E_{am}(q) = (\hbar^2/2I)a_m(q) + C_{\pm}$$

$$E_{bm}(q) = (\hbar^2/2I)b_m(q) + C_{\pm} \tag{11}$$

corresponding to two classes of Mathieu's functions

$$Y_{am}(z,q) = ce_m(z,q)$$

$$Y_{bm}(z,q) = se_m(z,q), \tag{12}$$

respectively. The parameter $q$ vanishing along with the rotational barrier $\varepsilon_{rot}$, Mathieu's eigenstates and eigenvalues convert to those of a *2-D free rotator*: $ce_0(z,0) = 1$, $ce_m(z,0) = \cos(mz)$, $se_m(z,0) = \sin(mz)$ and $a_m = m^2$, $b_m = m^2$.

The allowed rotational bands correspond to $q < 0$ (upper branch adiabatic energy) and $q > 0$ (lower branch adiabatic energy). These bands narrow as $|q|$ increases turning to single levels in the limit of too high barriers. Within an allowed energy band, Mathieu's eigenstates and eigenvalues are functions of the rotational wavenumber $k = n\pi$ where $0 \leq n \leq 1$ in the first Brillouin zone. For the particular case of a four-well potential energy ring at $Q = Q_0$, $n = \lambda/4$ where $\lambda = 1 \div 4$. The rotational energy bands are delineated as follows:

$(a_0, a_1), (b_1, b_2), (a_2, a_3), (b_3, b_4), ...$ for $q < 0$ (upper branch $\varepsilon_U$)

$(a_0, b_1), (a_1, b_2), (a_2, b_3), (a_3, b_4), ...$ for $q > 0$ (lower branch $\varepsilon_L$)

These bands are of alternating parities, even, odd, etc. in the increasing m order for the upper branch and are of mixed parity for the lower branch. Linear combinations of borderline states describe intra-band states, such as

$\sqrt{\frac{1}{2}}(ce_0+ce_1), \sqrt{\frac{1}{2}}(se_1+se_2), \sqrt{\frac{1}{2}}(ce_2+ce_3), \sqrt{\frac{1}{2}}(se_3+se_4),...q<0$ (upper branch $\varepsilon_U$)

$\sqrt{\frac{1}{2}}(ce_0+se_1), \sqrt{\frac{1}{2}}(ce_1+se_2), \sqrt{\frac{1}{2}}(ce_2+se_3), \sqrt{\frac{1}{2}}(ce_3+se_4),...q>0$ (lower branch $\varepsilon_L$)

etc. It should be stressed that the above assignment of allowed energy bands and band states to the upper and lower adiabatic branches holds good at $B_b > B_c$ and would be interchanged for $B_c > B_b$. The definition of a renormalized vibrational frequency $\omega_{ren}$ would also change accordingly.

Within an allowed band, Mathieu's eigenstates and eigenvalues should be attached a non-integer subscript, e.g. $ce_m(z,q)$, etc., where $0 \leq m \leq 1$. To specify the point Mathieu's eigenvalues $a_m(q)$ and $b_m(q)$ within an allowed band are expanded in power series in the rotational wave number $k = \pi m$. We exemplify the statement for the $(a_{n-1}, a_n)$ band. Starting from the lower borderline, we define intermediate eigenvalues by introducing a wave number:

$$a_{n-1+m}(q) = (n-1+m)^2 + \Delta a_{n-1+m}(q) \equiv m^2 + (n-1)(n-1+2m) + \Delta a_{n-1+m}(q)$$

where $0 \leq m \leq 1$ is a running number and the wave number is $0 \leq k \leq \pi$. These expansions are well known from the mathematical textbooks on Mathieu's eigenvalues which behave like free-rotator eigenvalues near $k = 0$, the band bottom. The remainders $\Delta a_n(q)$ contain renormalization terms that arise from the hindered interaction. We obtain a renormalized moment of inertia:

$$I_{n-1+m}^a(q) \equiv \hbar^2/\pi^2 (\partial^2 E_{n-1+m}^a(k)/\partial k^2)|_{k=0} = I / [1+\tfrac{1}{2}(\partial^2 \Delta a_{n-1+m}(q)/\partial m^2)]|_{m=0}$$

Accordingly, the rotational eigenstates near the bottom of a band should have the form of quasi-free eigenstates with renormalized inertial moments. In as much as $q$ increases as $I^2$, renormalized momenta should renormalize $q$ the other way round, so as quasi-free eigenstates should correspond to squeezed rotational bands for $\partial^2 \Delta a/\partial m^2 < 0$ leading to $I_{ren} > I$ and $q_{ren} > q$.

## 5. Optical conductivity

An approximate formula derived for the case of photon-assisted hopping of small polarons between neighboring cubic-symmetry sites reads [22]:

$$\sigma(\nu,T) = n_z N e^2 a_0^2 (\sqrt{\pi}/6\hbar)[t_e^2/k_B T \sqrt{(2\varepsilon_b k_B T)}] \exp(-\varepsilon_b / 2k_B T) \times$$

$$[\sinh(2\pi\hbar\nu / 2k_B T) / (2\pi\hbar\nu / 2k_B T)] \exp((2\pi\hbar\nu)^2/8\varepsilon_b k_B T), \quad k_B T \geq \tfrac{1}{2} \hbar\omega \quad (13)$$

Here $n_z$ is the coordination number, $\varepsilon_b = \tfrac{1}{2} \varepsilon_{gap} + \varepsilon_{JT}(1 - \varepsilon_{gap}/4\varepsilon_{JT})^2 - \tfrac{1}{2} \hbar\omega \approx \varepsilon_{JT} - \tfrac{1}{2} \times \hbar\omega$ ($\varepsilon_{gap} \ll 4\varepsilon_{JT}$) is the polaron binding energy, $t_e$ is the electron-transfer integral between nearest-neighbor sites, $\omega$ is the coupled phonon frequency, $\nu$ is the photon frequency, $a_0$ is the lattice constant, and $N$ is the polaron density. The tight-binding approximation to the electron energies is adopted throughout. The latter line of $\sigma(\nu,T)$ in equation (13) describes a nearly Gaussian band peaking at $2\pi \hbar\nu_{max} = 2\varepsilon_b$.

The optical conductivity $\sigma(\nu)$ relates to the optical absorption coefficient $\kappa(\nu)$ by

$$\sigma(\nu) = \nu n(\nu)\kappa(\nu) \tag{14}$$

where $n(\nu)$ is the refractive index of the medium. The optical absorption of a double well system in 1-D has been derived earlier in the harmonic approximation [23]. In view of the importance for our purpose we now extend the result so as to incorporate a temperature dependence. The assumed absorption coefficient reads:

$$\kappa(\nu) = [(2\pi)^3 N / 3c\,]\,\nu\,\textstyle\sum_{nn'}\rho_{Ln'}(T)|< \text{L}, n' \,|\, \mathbf{P} \,|\, \text{U}, n >|^2\,\delta(\varepsilon_{Un}-\varepsilon_{Ln'} -2\pi\eta\nu) \qquad (15)$$

where $N$ is the number of absorbing centers, L and U are the lower and upper branches of the adiabatic potential energy surface, n′ and n are the respective numbers of the vibronic energy levels thereat, $\rho_{Ln'}(T)$ is the thermal occupation factors in the lower state assumed to be in thermal equilibrium, and $\mathbf{P} = \sum e\mathbf{r}$ is the electric dipole associated with the optical transition.

### 5.1. Optical mixing dipole

Off-center displacements alike, the main electrostatic effects are expected to come from first-order electron-mode coupling. Retaining the first-order terms, the adiabatic eigenstates corresponding to $\varepsilon_{U/L}(Q)$ in (2) are [17]:

$$|\,\mathbf{r},\mathbf{Q}>_{U,L} = \mu\sqrt{\tfrac{1}{2}}\{[(\textstyle\sum_{i=x,y,z}4G_i^2Q_i^2 + \varepsilon_{\alpha\beta}^2)^{1/2}\mu\,\varepsilon_{\alpha\beta}]\,|\,a_{1g}> \mu\,\textstyle\sum_{i=x,y,z}2G_iQ_i\,|\,t_{1u\,i}>\}\,/$$

$$\sqrt{\{[\textstyle\sum_{i=x,y,z}4G_i^2Q_i^2 + \varepsilon_{\alpha\beta}^2]^{1/2}\,[(\textstyle\sum_{i=x,y,z}4G_i^2Q_i^2 + \varepsilon_{\alpha\beta}^2)^{1/2}\mu\,\varepsilon_{\alpha\beta}]\}} \qquad (16)$$

These are mixed states coinciding with the basis states at $Q_i = 0$, the non-mixing (central) configuration: $|\,\mathbf{r},\mathbf{0}>_U = \sum_I |\,t_{1u\,i}>$, $|\,\mathbf{r},\mathbf{0}>_L = |\,a_{1g}>$. At larger $Q_i \ll \varepsilon_{\alpha\beta}/2G_i$, $|\,\mathbf{r},\mathbf{Q}>_U = \sqrt{\tfrac{1}{2}}(\,|\,a_{1g}> -\,\sum_i |\,t_{1u\,i}>)$, $|\,\mathbf{r},\mathbf{Q}>_L = \sqrt{\tfrac{1}{2}}(\,|\,a_{1g}> +\,\sum_i |\,t_{1u\,i}>)$, the anti-symmetric and symmetric combinations of the basis eigenstates, respectively.

The adiabatic mixing dipole then is:

$$\mathbf{P}_{ad\,UL}(\mathbf{Q}) = {}_U<\mathbf{r};\mathbf{Q}\,|\,e\mathbf{r}\,|\,\mathbf{r};\mathbf{Q}>_L =$$

$$-\tfrac{1}{2}\textstyle\sum_{i=x,y,z}2G_iQ_i\,\big[(<a_{1g}|\,e\mathbf{r}\,|\,t_{1u\,i}> - <t_{1u\,i}|\,e\mathbf{r}\,|\,a_{1g}>) -$$

$$\varepsilon_{\alpha\beta}\,(<a_{1g}|\,e\mathbf{r}\,|\,t_{1u\,i}> + <t_{1u\,i}|\,e\mathbf{r}\,|\,a_{1g}>)(\textstyle\sum_{i=x,y,z}4G_i^2Q_i^2 + \varepsilon_{\alpha\beta}^2)^{-1/2}\,\big]\,/$$

$$(\textstyle\sum_{i=x,y,z}4G_i^2Q_i^2 + \varepsilon_{\alpha\beta}^2)^{1/2} \qquad (17)$$

This electric dipole controls the optical transitions between the lower and upper adiabatic states. It would be informative to assess its magnitude at the off-center radius $Q_0$, e.g. for cubic symmetry:

$$\mathbf{P}_{ad\,UL}(Q_0) = \sqrt{3}<a_{1g}|\,\mathbf{r}\,|\,t_{1ui}>(\varepsilon_{\alpha\beta}/4\varepsilon_{JT})\,[1-(\varepsilon_{\alpha\beta}/4\varepsilon_{JT})^2]^{1/2}$$

$$= \sqrt{3}<a_{1g}|\,\mathbf{r}\,|\,t_{1ui}>(\varepsilon_{\alpha\beta}\,K^2/2G^3)\,eQ_0 < eQ_0 \qquad (18)$$

To adiabatic approximation, the vibronic dipolar matrix element is given by:

$$\mathbf{P}_{vib\,UL} \equiv <\text{L},n'\,|\,\mathbf{P}\,|\,\text{U},n> = <n',Q\,|<\text{L},Q\,|\,\mathbf{P}\,|\,\text{U},Q>|\,n,Q>$$

$$= <n',Q\,|\,\mathbf{P}_{ad\,UL}(\mathbf{Q})\,|\,n,Q> \qquad (19)$$

in so far as the total Born-Oppenheimer wave function is factorized out of electronic and vibrational wave functions. The middle term $\mathbf{P}_{ad\ UL}(\mathbf{Q})$ has been calculated above using the adiabatic wavefunctions. The complete matrix element $\mathbf{P}_{vib\ UL}$ depends on the vibronic wave functions which are not always known exactly, due to the anharmonicity of the problem.

### 5.2. Harmonic approximation to 1D flip-flops

In 1-D, the adiabatic wave functions corresponding to $\varepsilon_{U/L}(Q)$ read simply [17]:

$| L,Q \rangle = 2^{-\frac{1}{2}} [\cos(\phi/2) + \sin(\phi/2)] | a_{1g} \rangle + 2^{-\frac{1}{2}} [\cos(\phi/2) - \sin(\phi/2)] \sum_i | t_{1ui} \rangle$

$| U,Q \rangle = -2^{-\frac{1}{2}} [\cos(\phi/2) - \sin(\phi/2)] | a_{1g} \rangle + 2^{-\frac{1}{2}} [\cos(\phi/2) + \sin(\phi/2)] \sum_i | t_{1ui} \rangle$

where

$\phi(Q) = \tan^{-1}(\varepsilon_{\alpha\beta}/2GQ)$

At $Q = 0$ (central configuration), $\phi = \frac{1}{2}\pi$: $| L,Q \rangle = | a_{1g} \rangle$, $| U,Q \rangle = \sum_i | t_{1ui} \rangle$; at large $Q_0 \gg \varepsilon_{\alpha\beta}/2G$, $\phi = 0$: $|L,Q\rangle = 2^{-\frac{1}{2}}(\sum_i | t_{1ui} \rangle + | a_{1g} \rangle)$, $| U,Q \rangle = 2^{-\frac{1}{2}}(\sum_i | t_{1ui} \rangle - | a_{1g} \rangle)$, the symmetric and antisymmetric combinations of the basis eigenstates, respectively. The following adiabatic mixing dipole is obtained:

$\mathbf{P}_{ad\ UL}(Q) \equiv \langle L,Q | \mathbf{P} | U,Q \rangle = \sin\phi(Q) \langle a_{1g} | \mathbf{P} | t_{1u} \rangle$

$\qquad = \{\varepsilon_{\alpha\beta}/\sqrt{[(2GQ)^2 + \varepsilon_{\alpha\beta}^2]}\} \langle a_{1g} | \mathbf{P} | t_{1u} \rangle$ \qquad (20)

The early work employed harmonic-oscillator wave functions at renormalized vibrational frequencies to calculate the vibronic dipolar matrix elements:

$\langle 0,Q | \mathbf{P}_{ad\ UL}(Q) | n,Q \rangle = \sqrt{\{(\alpha_U \alpha_L / 2^{n+1} n! \pi)[1 \pm \exp(-\xi_0)]\}} \exp(-\xi_0^2) \times$

$_{-\infty}\int^{+\infty} \mathbf{P}_{ad\ UL}(Q) H_n(\xi_0) \exp(-(\xi_U^2 + \xi_L^2)/2)(\exp(\xi_U \xi_L) \pm \exp(-\xi_U \xi_L)) dQ$ \qquad (21)

where $\omega_U = \omega\sqrt{(1+\eta^{-1})}$ and $\omega_L = \omega\sqrt{(1-\eta^2)}$ for $\eta = \varepsilon_{\alpha\beta}/4\varepsilon_{JT}$ are the renormalized frequencies at well bottom of $\varepsilon_U(Q)$ and $\varepsilon_L(Q)$, respectively, $\alpha_U = \sqrt{(M\omega_U/\hbar)}$, $\alpha_L = \sqrt{(M\omega_L/\hbar)}$, $\xi_L = \alpha_L Q$, $\xi_U = \alpha_U Q$, $\xi_0 = \alpha_L Q_0$. The optical transition energy is:

$2\pi \hbar \nu_n = \varepsilon_{JT}(1+\eta^2) + \frac{1}{2} \varepsilon_{\alpha\beta} + \frac{1}{2} \hbar(\omega_U - \omega_L) + n\hbar\omega_U \pm \frac{1}{2}\Delta$ \qquad (22)

where $\Delta = \varepsilon_{\alpha\beta} \exp(-2\varepsilon_{JT}/\hbar\omega_L)$ is the ground-state tunneling splitting, $\pm$ refer to the symmetric and antisymmetric combinations of the lowest vibronic eigenstates of the two wells, respectively.

Numerical values of the fitting parameters entering in equations above are listed in Table I. They are in concert with parameters from literature data.

### 5.3. Optical conductivity of 2D rotors

The optical absorption associated with transitions at off-center rotators can be derived proportional to the absolute of the electrostatic dipole $\mathbf{P}_{ad\ UL}(Q)$ taken between initial and final Mathieu states [23]. At low temperature, only the lowest rotational band being populated, all optical transitions start from levels in that band:

$$\kappa(\nu) \propto |\sum_{nm} < ce_{0+m}(z,q) | \mathbf{P}_{ad\ UL}(Q_l) | se_{n+m}(z,q) > |^2, \quad (0 < m < 1, n=1,2,3,...) \quad (23)$$

This equation conserves the rotational momentum $k = \pi m$, as required by the selection rules for optical transitions.

For deriving an absorption coefficient, rewriting the mixing dipole $\mathbf{P}_{UL}(Q)$ is advisable. For the dipole associated with a rigid 3-D off-center rotor winding along the spherical brim at $|\mathbf{Q}| = Q_0$, we get:

$$\mathbf{P}_{ad\ UL}(\mathbf{Q}_0)^{3D} = {}_U< \mathbf{r};\mathbf{Q}_0 | e\mathbf{r} | \mathbf{r};\mathbf{Q}_0 >_L =$$

$$\varepsilon_{\alpha\beta} \sum_{i=x,y,z} 2G_i Q_{i\,0} (<a_{1g}|e\mathbf{r}|t_{1ui}>) / (4G^2 \sum_{i=x,y,z} Q_{i\,0}^2 + \varepsilon_{\alpha\beta}^2) =$$

$$(2GQ_0/4\varepsilon_{JT})(\varepsilon_{\alpha\beta}/4\varepsilon_{JT}) \sum_{i=x,y,z} (Q_i/Q_0) <a_{1g}|e\mathbf{r}|t_{1ui}> = \quad (24)$$

$$(\varepsilon_{\alpha\beta}/4\varepsilon_{JT})[1-(\varepsilon_{\alpha\beta}/4\varepsilon_{JT})^2]^{1/2} \times$$

$$[\sin\theta\cos\varphi <a_{1g}|e\mathbf{r}|t_{1ux}> + \sin\theta\sin\varphi <a_{1g}|e\mathbf{r}|t_{1uy}> + \cos\theta <a_{1g}|e\mathbf{r}|t_{1uz}>]$$

For a 2-D rotor winding along the circular brim at $\theta = \frac{1}{2}\pi$ with $z = 2\varphi$,

$$\mathbf{P}_{UL}(\mathbf{Q}_0)^{2D} = (\varepsilon_{\alpha\beta}/4\varepsilon_{JT})[1-(\varepsilon_{\alpha\beta}/4\varepsilon_{JT})^2]^{1/2} \times$$

$$[\cos(\tfrac{1}{2}z) <a_{1g}|e\mathbf{r}|t_{1ux}> + \sin(\tfrac{1}{2}z) <a_{1g}|e\mathbf{r}|t_{1uy}>] \quad (25)$$

The vibronic matrix elements of $\mathbf{P}_{ad\ UL}(Q_0)$ will have to be calculated between initial and final intraband states which we define as follows, respectively:

$$ce_{0+m}(z,q) = [m^2 + (1-m)^2]^{-1/2} [(1-m) ce_0(z,q) + m\, ce_1(z,q)]$$

$$(ce_{n-1}, se_n)_{+m}(z,q) = [m^2 + (1-m)^2]^{-1/2}[(1-m) ce_{n-1}(z,q) + m\, se_n(z,q)] \quad (26)$$

with corresponding eigenvalues

$$a_{0+m}(q) \equiv (2I/\hbar^2) <ce_{0+m}(z,q)|H|ce_{0+m}(z,q)>$$

$$= [m^2 + (1-m)^2]^{-1} [(1-m)^2 a_0(q) + m^2 a_1(q)],$$

$$(a_{n-1}, b_n)_{+m}(q) \equiv (2I/\hbar^2) < (ce_{n-1}, se_n)_{+m}(z,q)|H|(ce_{n-1}, se_n)_{+m}(z,q) >$$

$$= [m^2 + (1-m)^2]^{-1} [(1-m)^2 a_{n-1}(q) + m^2 b_n(q)] \quad (27)$$

It is easy to verify that at small rotational momenta $k = \pi m$ the intraband eigenenergies

are linear in k², as they should:

$a_{0+m}(q) \sim a_{0'}(q) + (k/\pi)^2 a_1(q)$

$(a_{n-1}, b_n)_{+m}(q) \sim a_{n-1}(q) + (k/\pi)^2 b_n(q)$  (28)

The assumed absorption coefficient will then read:

$\kappa(\nu,q) = [(2\pi)^3 N / 3c] \sum_{nm} \nu_{nm} \rho_{Lm}(T) \times$

$|<ce_{0+m}(z,-q) | \mathbf{P}_{ad\ UL}(\mathbf{Q}) | (ce_{n-1}, se_n)_{+m}(z,q)>|^2 \delta(\varepsilon_{U\ n+m} - \varepsilon_{L\ 0+m} - 2\pi\hbar\nu_{nm})$  (29)

where

$\varepsilon_{U\ n+m} = (\hbar^2/2I)(a_{n-1}, b_n)_{+m}(+q)$

$\varepsilon_{L\ 0+m} = (\hbar^2/2I) a_{0+m}(-q)$

where $N$ is the number of absorbing centers, L(−q) and U(+q) are the lower and upper branches of the adiabatic potential energy surface, n and m are the respective numbers of the vibronic energy levels, $\rho_{Lm}(T)$ is the thermal occupation factor in the lowest state assumed to be in thermal equilibrium, and $\mathbf{P}_{ad\ UL}(\mathbf{Q})$ is the adiabatic electric dipole associated with the optical transition.

Performing the delta function operation in (29) turns the spectral function into an envelope of absorptive lines at discrete photon frequencies according to

$\kappa(\nu_{nm},q) = [(2\pi)^3 N /3c]\nu_{nm}\rho_{Lm}(T)|<ce_{0+m}(z,-q) | \mathbf{P}_{adUL}(\mathbf{Q}) | (ce_{n-1}, se_n)_{+m}(z,+q)>|^2$  (30)

where the frequencies are

$\nu_{nm} = (\varepsilon_{U\ n+m} - \varepsilon_{L\ 0+m}) / 2\pi\hbar = (\hbar/4\pi I) \{(a_{n-1}, b_n)_{+m}(+q) - a_{0+m}(-q)\} + \Delta C_\pm / 2\pi\hbar$

$= (\hbar/4\pi I)[m^2 + (1-m)^2]^{-1} \{(1-m)^2 [a_{n-1}(+q) - a_0(-q)] +$

$m^2 [b_n(+q) - a_1(-q)]\} + \Delta C_\pm / 2\pi\hbar$  (31)

Inserting (26) into (30) we rewrite (30) further

$\kappa(\nu_{nm},q) = [(2\pi)^3 N /3c] \nu_{nm} \rho_{Lm}(T) [m^2 + (1-m)^2]^{-2} \times$

$|<[(1-m)ce_0(z,-q) + m\ ce_1(z,-q)]|\mathbf{P}_{ad\ UL}(\mathbf{Q})|(1-m)ce_{n-1}(z,+q) + m\ se_n(z,+q)>|^2$

$= [(2\pi)^3 N /3c]\nu_{nm}\rho_{Lm}(T)[m^2 + (1-m)^2]^{-2} \times$

$|m(1-m)<ce_0(z,-q) | \mathbf{P}_{ad\ UL}(\mathbf{Q}) | se_n(z,+q)> + m^2 <ce_1(z,-q) |\mathbf{P}_{ad\ UL}(\mathbf{Q})| se_n(z,+q)> +$

$(1-m)^2<ce_0(z,-q)|\mathbf{P}_{ad\ UL}(\mathbf{Q})|ce_{n-1}(z,+q)> +$

$$m(1-m) |\langle ce_1(z,-q) | \mathbf{P}_{ad\ UL}(\mathbf{Q}) | ce_{n-1}(z,+q) \rangle|^2 \quad (32)$$

The last two terms of the sum in (32) do not vanish, since the rotational dipole moment in (25) is not definite-parity. The finite third term therein makes the absorption cross-section non-vanishing at $m = 0$, unlike the transitions controlled by the momentum-conserving selection rule. At intermediate $0 < m < 1$ all lowest band levels contribute to the absorption. At $m \to 1$, transitions starting from the top of the lowest $(a_0, a_1)$ band will predominate in the optical absorption. Equation (32) displays a peak between $0 < m < 1$, as in earlier harmonic-oscillator based theories. We also see that the rotational barrier $\varepsilon_{rot}$ enters into equation (32) via q by virtue of equation (10) in that $|q| = 0$ yields a free rotation, $|q| \sim 1$ brings a hindered rotation, and $|q| \gg 1$ leads to immobilization.

### 5.4. Numerical calculations

#### 5.4.1. Working conditions

In as much as the exact eigenfunctions of a hindered 3-D rotator are not available, we focused on the optical properties of planar rotators described by Mathieu functions. By partitioning a rotational band into equidistant intraband states as in (27), the matrix element in (29) at given *n* is made simpler. At first, we derived a spectral envelope function for the absorption curve discarding the intraband details. In so far as the number of off-center rotational sites in-plane is 4, we have e.g. $m = 0, 0.25, 0.5, 0.75, 1$. For deriving an envelope function, the discrete photon frequencies obtain from (31) at the edge value of $m = 1$:

$$\nu_n = (\varepsilon_{U\ n+1} - \varepsilon_{L\ 1}) / 2\pi \hbar = (\hbar / 4\pi I)[b_{n+1}(+q) - a_1(-q)] + \Delta C_\pm / 2\pi \hbar \quad (n=0,1,2,\dots) \quad (33)$$

and we get for that function

$$\kappa(\nu, q) \propto \sum_n \nu_n |\langle ce_1(z,-q) | \mathbf{P}_{ad\ UL}(\mathbf{Q}) | se_n(z,+q) \rangle|^2 \delta(\varepsilon_{U\ n+1} - \varepsilon_{L\ 1} - 2\pi \hbar \nu_n) \quad (34)$$

or, equivalently,

$$\kappa(\nu_n, q) \propto \nu_n |\langle ce_1(z,-q) | \mathbf{P}_{ad\ UL}(\mathbf{Q}) | se_n(z,+q) \rangle|^2 \quad (35)$$

provided $\sum_m \nu_{nm} \rho_{Lm}(T) = \nu_n \sum_m \rho_{Lm}(T) = \nu_n$. In equations (31) & (33) $\Delta C_\pm = C_+ - C_-$ where $C_+$ and $C_-$ are defined by equations (9). Equation (35) is the absorption cross-section by transitions from the energy level at $a_1(q)$ to a series of edge levels at $b_n(q)$ disregarding completely any transitions to the intraband levels. Results of numerical calculations based on the foregoing analysis with parameters taken from Table I are presented in Figure 1 (a)-(c).

Table I
Empirical and calculated point-ion parameters

| | | | | | | | |
|---|---|---|---|---|---|---|---|
| | | | | | | | |

| Host | Cavity radius $r_0$ (Å) | Madelung energy $V_M$ (eV) | Dielectric constant $\kappa_0$ $\kappa_p$ $\kappa_s$ | Stiffness K (eV/Å$^2$) | Gap $\varepsilon_{\alpha\beta}$ (eV) | Coupling constant G (eV/Å) | Coupling constant $B_c$ (eV/Å$^3$) | Coupling constant $B_b$ (eV/Å$^3$) |
|---|---|---|---|---|---|---|---|---|
| KCl:Li | 3.14 | 7.9 | 2.1  4  4.8 | 4.38 | 1.41 | 2.25 | -0.18 | 0.36 |
| LSCO | - | - | -  -  - | 7.50 | 0.9 | 2.45 | - | - |
| YBCO | - | - | -  -  - | 10.46 | 1.5 | 3.63 | - | - |

| Host | JT energy $\varepsilon_{JT}$ (eV) | Phonon frequency $\omega_{renI}$ (meV) 3D 2D 1D | Phonon frequency $\omega_{renII}$ (meV) | Rotational barrier $\varepsilon_{BII}$ (meV) | Modemass M ($M_{anion}$ u.) 3D 2D 1D | Off-site radius $Q_0$ (Å) | Roton energy $\hbar^2/2I$ (meV) 3D  2D |
|---|---|---|---|---|---|---|---|
| KCl:Li | 0.58 | 5.0  6.1 | 8.2 | 9.9 | 1.5  1  - | 0.32 | .43  .65 |
| LSCO | 0.40 | 5.2 | | 77 | 0.5 | 0.27 | |
| YBCO | 0.63 | 5.9 | | 103 | 0.5 | 0.28 | |

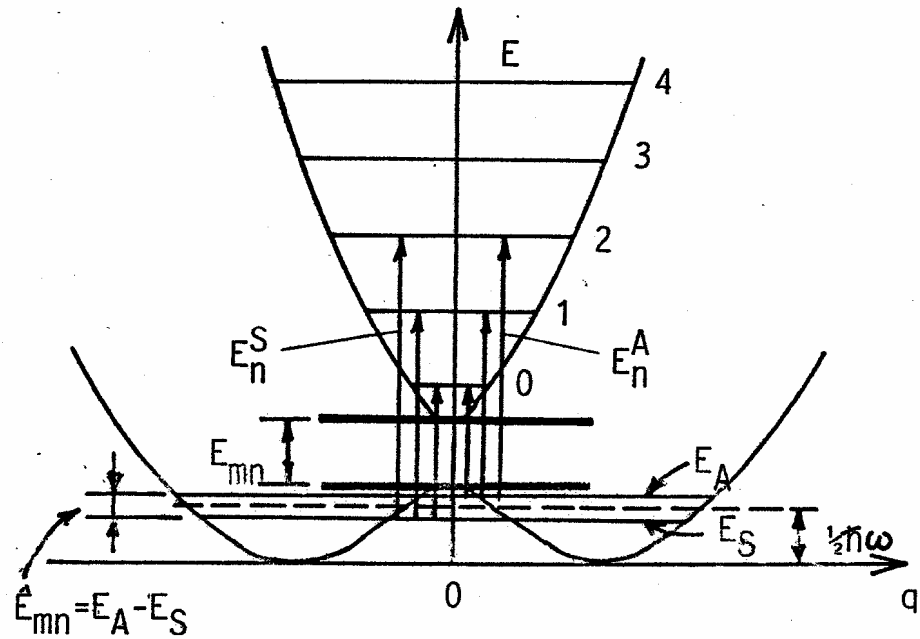

Figure 1 (a)

Figure 1 (a). Illustrating vibronic energies and optical transitions at a 1D double well.

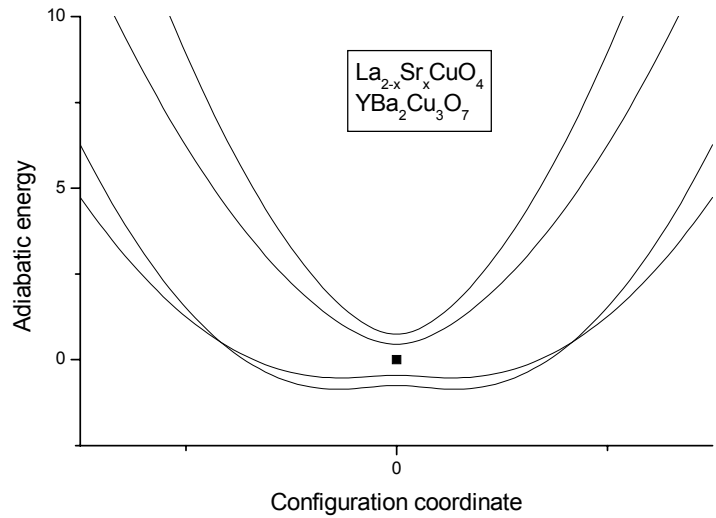

Figure 1 (b)

Figure 1 (b). Showing the derived double-well potentials of linear off-center O(A) host ions in $La_{2-x}Sr_xCuO_4$ (LSCO) and $YBa_2Cu_3O_7$ (YBCO).

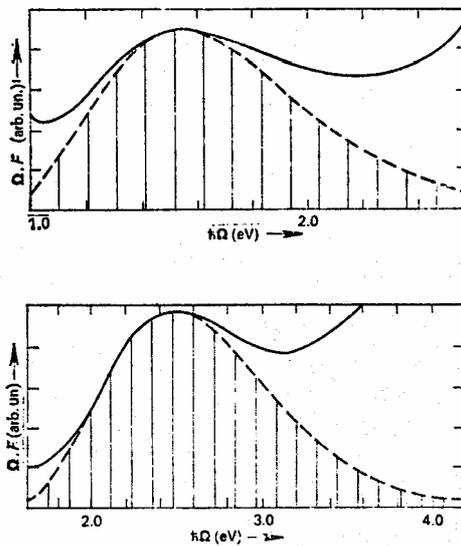

Figure 1 (c)

Figure 1 (c). Showing the optical absorption spectra associated with off-center apical oxygen ions in $La_{2-x}Sr_xCuO_4$ (LSCO) and $YBa_2Cu_3O_7$ (YBCO) upside down. The theoretical envelope spectra calculated by the tabulated parameters (broken line) are compared with the experimental spectra (solid line), after Reference [33]. Note that the linear off-center ions reorientate via 1-D flip-flops along a-, b-, or c-axes in high-$T_c$ superconductors.

### 5.4.2. Numerical parameters

For carrying out numerical calculations, values for the Mexican hat parameters will have to be specified. We used estimated values for KCl, as exemplified in Table I. For a brief retrospective outlook, we assume a spherical-well point-ion potential (Devonshire type) at the impurity site ($r_0$ - cavity radius, $\alpha_M$ - Madelung's constant, $V_M = \alpha_M e^2/r_0$ - Madelung's energy)

$$U(r,Q_l) = (\alpha_M e^2/r_0)\{r_0 /\sqrt{[(x_0 - Q_x)^2+(y_0 - Q_y)^2+(z_0 - Q_z)^2]}\} \tag{36}$$

creating an electric field $\mathbf{F}(r_0,Q_l) = -\text{grad}_\mathbf{Q} U(r_0,Q_l)$ that couples to the dipole $e\mathbf{r}$ mixing the electronic states at the defect site. The coupling energy $V = -e\mathbf{r}.\mathbf{F}$ is

$$V(r_0,Q_l) = -\alpha_M e\{ex(x_0+Q_x) + ey(y_0+Q_y) + ez(z_0+Q_z)\}/$$
$$[(x_0+Q_x)^2 + (y_0+Q_y)^2 + (z_0+Q_z)^2]^{3/2} \tag{37}$$

Now, we define 1st- and 3rd- order coupling operators according to

$$b_x = \partial V(r,Q)/\partial Q_x |_{Ql=0} = \alpha_M e^2/r_0^2$$

$$d_{xyz} = \partial^3 V(r,Q)/\partial Q_x \partial Q_y \partial Q_z |_{Ql=0} = 74xy_0z_0/r_0^7$$

$$d_{xxz} = \partial^3 V(r,Q)/\partial Q_x^2 \partial Q_z |_{Ql=0} = -6z_0(rr_0 + 11xx_0)/r_0^7$$

$$d_{xxx} = \partial^3 V(r,Q)/\partial Q_x^3 |_{Ql=0} = 6x(3r_0^2 - 11x_0^2)/r_0^7, \tag{38}$$

etc. These operators are to be integrated between the $|a_{1g}>$ and $|t_{1u}>$ electronic states to obtain the respective mixing constants. This will be reported elsewhere. For a cubic alkali halide $\alpha_M = 1.74$. The KCl data are summarized in Table I. We set $M_{3D} = (3/2)M_{Cl}$, $M_{2D} = M_{cl}$ and $K = M_{3D}\omega_{3D}^2 = M_{2D}\omega_{2D}^2$. We use KCl IR data $\omega_{3D} = \omega_{IR\,3D}$ (40 cm$^{-1}$), $\omega_{2D} = \omega_{IR\,2D}$ (49 cm$^{-1}$) to compute K and thereby $\varepsilon_{JT} = b^2/2K$. We have also got $\varepsilon_{gap}$ and $\varepsilon_{JT}$ from Li$^+$ cluster calculations. Using the tabulated values, we calculate $C_+ = 1.5273$ eV, $C_- = -0.7957$ eV and thereby $\Delta C_\pm = 2.323$ eV.

### 5.4.3. Calculations

To begin with, calculations using (35) are devised as 3-dimensional graphics in the (v,q) plane. We remind that barriers becoming too high in the limit of large q, the off-center ion is too slow to rotate effectively along the sombrero brim, though it remains localized somewhere along the brim periphery. As q gets smaller, the hindering becomes less formidable converting the motion to a genuine rotation at not too large q. Finally, as q tends to vanish, the ion approaches the limit of quasi-free rotation. As to the dipoles entering in (32) to (35) we reasonably set $< a_{1g} | e\mathbf{r} | t_{1ux} > = < a_{1g} | e\mathbf{r} | t_{1u} >$ so as to factorize their contribution out of the absorption cross-section integrals.

Extensions based on the complete equation (29) so as to incorporate the temperature dependence of the absorption spectra are also devised. We remind that the difference between formulae (29) and (32) is in the latter one degrading the bandwidth which is actually non-vanishing and increasing with the band suffix n. Indeed, with $I \sim 10^{-40}$ g.cm$^2$ leading to $\hbar^2/2I \sim 3$ meV we get the bandwidths increasing roughly as $(\hbar^2/2I)[m^2-(m-1)^2] = 3, 9, 15, 21$, etc. in meV. Nevertheless, both formulas (29) and (32) should give convergent results for the shape of the $(a_1 \rightarrow b_n)$ band-edge transitions. We next construct two more 2D graphics for the absorption cross section $k(\nu,T,q)$ in the $(\nu,T)$ and $(\nu,q)$ planes, respectively, using the complete formula (29).

## 6. Experimental data and analysis

### 6.1. The FIR band at 3-D Li$^+$ impurity off-site rotators in alkali halides

Looking for an optical conductivity so as to check the theoretical models, the Li$^+$ impurity in alkali halides provides a most favorable object because of its clear far-infra- red (FIR) absorption bands.

Indeed, a clear optical signature of the off-center Li$^+$ impurity seems to appear as a FIR absorption band centered at $\sim 40$ cm$^{-1}$ in KCl [24]. The optical band shows an anomalous Li-isotope effect. We suggest that Li$^+$ in uncolored KCl rotates as part of a massive frame composed of three Cl$^-$–Cl$^-$ pairs or $T_{1u}$ oscillators, dragging it to <111> sites if the frame-ion coupling is strong enough [25]. This frame should experience little mass change on Li$^+$ isotope substitution. In as much as the $T_{1u}$ vibrational frequencies are in the TO range $\sim 100$ cm$^{-1}$, the drop to $\sim 40$ cm$^{-1}$ implies renormalization due to the coupling. In so far as Li$^+$ has been dragged off-center, the coupling is strong ($4E_{JT} > E_{gu}$) and the renormalized frequency reads

$$\omega_{ren\ sc} = \omega_{bare}\{1 - (E_{gu}/4E_{JT})^2\}^{1/2} \qquad (39)$$

The assignment of bare (100 cm$^{-1}$) and renormalized (40 cm$^{-1}$) mode frequencies yields $4E_{JT}/E_{gu} = 1.091$. From cluster calculations, we have $E_{gu} = 4$ eV whereby $E_{JT} = 1.091$ eV. Using a $(3/2)M_{Cl}$ reduced oscillator mass, we compute a spring constant $K = 1.967$ eV/Å$^2$ and thereby a vibronic mixing constant $G = (2E_{JT}K)^{1/2} = 2.072$ eV/Å. The off-center displacement $Q_0 = \{(2E_{JT}/K)[1 - (E_{gu}/4E_{JT})^2]\}^{1/2}$ is estimated at 0.42 Å, the on-center→off-center barrier height $E_b = E_{JT}(1 - E_{gu}/4E_{JT})^2$ is 8 meV so that there are two underbarrier levels at 2.5 meV and 7.5 meV, respectively, at the renormalized vibrational frequency 40 cm$^{-1}$ = 5 meV. The FIR optical absorption and Raman active bands associated with lithium in uncolored KCl are seen in Figure 2 (a).

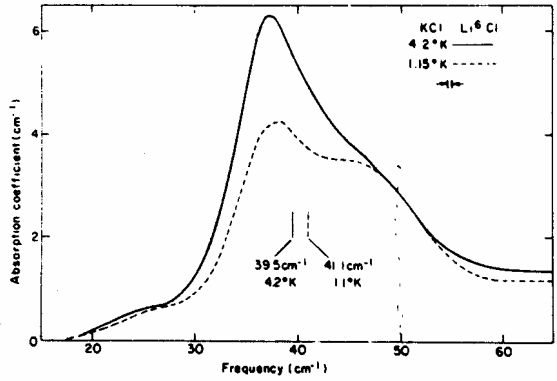

Figure 2 (a)

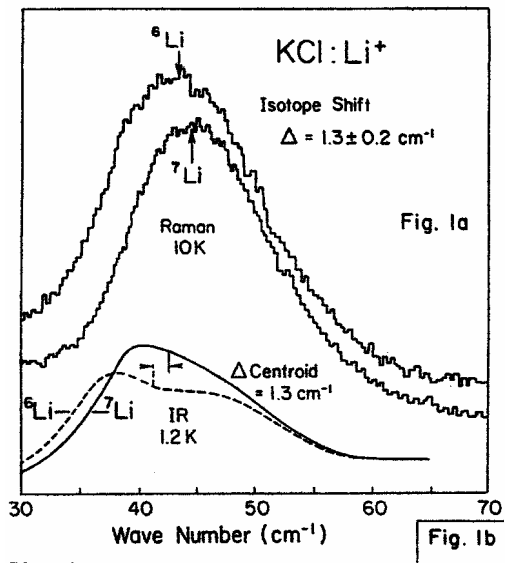

Figure 2 (b)

Figure 2. The FIR optical absorption bands (a) and the Raman active bands (b) in the vicinity of 40 cm$^{-1}$ characteristic of the Li$^+$ impurity in uncolored KCl, from Refs. [25] and [29], respectively. The small anomalous isotope shift on $^7$Li to $^6$Li substitution can also be seen. Both bands shift to about 47 cm$^{-1}$ when the samples are colored and underwent optical and thermal treatment so as to precipitate F centers at sites nearest neighboring the impurity and form F$_A$ centers. Physically, the frequency hardening is due to the immobilization of one of the three Cl$^-$–Li$^+$–Cl$^-$ vibrating pairs coupled to Li$^+$ on reducing dimensionality from 3-D to 2-D.

The former has been the subject of considerable debate in the literature [24–28]. This is a doublet possibly triplet band with components at ~ 25 cm$^{-1}$, 38 cm$^{-1}$ and 43 cm$^{-1}$. Although the former two might be associated with lithium aggregates, the latter one seems inherent to the 3-D rotator by isolated off-center Li$^+$ ions. Their eigenstates are largely unknown at present. Whenever an off-center displacement occurs, the inversion symmetry at the impurity centered cluster is broken and the renormalized frequency appears in both the IR absorption and Raman scattering spectra.

Li$^+$ does not go off-center in all the alkali halide hosts, certainly not in NaCl. However the frame-ion coupling is still there, even though weak, for the FIR band again appears at a renormalized frequency

$$\omega_{\text{ren wc}} = \omega_{\text{bare}} (1 - 4E_{\text{JT}} / E_{\text{gu}})^{1/2} \qquad (40)$$

This equation is the basis for analyzing weak-coupling ($4E_{\text{JT}} < E_{\text{gu}}$) data.

### 6.2. The FIR band at 2-D Li$^+$ impurity off-site rotators in colored alkali halides

The most profound effect on Li$^+$ dynamics of the reduced 3D → 2D dimensionality is the increase of the renormalized frequency $\omega_{\text{ren}}$ in proportion of $(3/2)^{1/2}$. This results from the drop in reduced mass of the $T_{1u}$ oscillator from $(3/2)M_{Cl}$ to $(2/2)M_{Cl}$ as a Cl$^-$–Cl$^-$ vibrating pair is frozen when one of its ion components is replaced by an F center [25,26]. (This pair will largely soften its frequency due to reduced stiffness.) To first approximation the proximity of an F center will not alter the spring constant K of surviving pairs. The reduced-D effect is observed through a small hardening of the FIR & Raman bands from 43 cm$^{-1}$ to 47 cm$^{-1}$, as one goes from KCl:Li$^+$ to KCl:F$_A$(Li$^+$) [29]. We conclude that the proximity of an F center does not seem to alter too much the off-site behavior of the Li impurity. The FIR and Raman active bands in colored KCl:Li$^+$ are shown in Figure 2 (b).

Both lithium bands exhibit a small anomalous isotope shift on $^7$Li to $^6$Li substitution, as seen in Figure 2 (a), which lends support to the Cl$^-$ pair $T_{1u}$ assignment to the rotating mode coupling [25,26]. As Li$^+$ goes off-center, the site symmetry is transfigured from cubic O$_h$ to axial C$_{4v}$, while the mode nomenclature changes from A$_{1g}$ to A$_1$. With the breakdown of inversion symmetry the infrared active modes become Raman active too.

### 6.3. O(A)-Cu(P)-O(A) axial bonds in the La$_{2-x}$Sr$_x$CuO$_4$ family

The optical mixing dipole of equation (20) may be compared with the off-center dipole under the harmonic approximation in 1-D which reads [30]:

$$\mathbf{P}_{\text{ad LL/UU}}(Q) \equiv < L,Q \mid \mathbf{P} \mid L,Q > = \cos\phi(Q) < a_{1g} \mid \mathbf{P} \mid t_{1u} >$$

$$= \pm\{(2GQ)/\sqrt{[(2GQ)^2 + \varepsilon_{\alpha\beta}^2]}\} < a_{1g} \mid \mathbf{P} \mid t_{1u} > \qquad (41)$$

The + (–) signs in equation (41) relate to the lower (upper) branches of the vibronic (adiabatic) energy. We see that the off-center dipole is inherent to both branches even though it appears with opposite signs up or down. In the latter case, the off-center dipole is vanishing at the nonmixing configuration ($Q = 0$) and increases to saturate as one goes away from it. From equation (20), the optical mixing dipole is complementary to the off-center dipole, for the former one goes to its highest as $Q \to 0$ where the latter one vanishes.

It will also be informative to derive the dc-conductivity arising from an array of small polarons in 1-D. From equation (13) setting $v = 0$ we get

$$\sigma(0,T) = n_z N e^2 a_0^2 (\sqrt{\pi}/6\text{h})[t_e^2/k_B T \sqrt{(2\varepsilon_b k_B T)}]\exp(-\varepsilon_b/2k_B T) \qquad (42)$$

We see the dc-conductivity of small polarons to be thermally activated by the polaron binding energy. The temperature-dependent pre-exponential factor is inherent of a polaron gas obeying the Boltzmann-tail statistics. The overall dc-conductivity tends to nil as the temperature tends to the absolute zero, unlike the temperature behavior of small polarons along a copper-oxygen bond in high-$T_c$ superconductors which have instead been reported to exhibit a divergent conductivity as the sample is cooled down at the lowest temperatures [31]. From equation (13) using (42) we get for the optical conductivity

$$\sigma(\nu,T) = \sigma(0,T)[\sinh(2\pi\eta\nu/2k_B T)/(2\pi\eta\nu/2k_B T)]\exp((2\pi\eta\nu)^2/8\varepsilon_b k_B T), \quad k_B T \geq \tfrac{1}{2}\eta\omega \qquad (43)$$

Equation (43) may be regarded as a generalized form of an optical conductivity in which the frequency-dependent part is more or less tradtional, while the off-center effects are accounted for by terms entering into the dc-factor. Among these terms is the hopping energy $t_e$ which is essentially modified as the off-center entity is smeared within the off-site volume. Due to smearing, the nearest-neighbor wave-function overlap will be enhanced leading to an increased hopping energy. Because of the lack of more accurate data, it is hard at this time pointing to other off-center related effects on the dc-conductivity. Some useful suggestions of this kind can be found in a recent publication [11].

Perhaps, the dc-conductivity problem has been addressed more cautiously albeit not exhaustively in a subsequent publication [15]. There the dc-conductivity along a Cu-O bond in a single-plane high-Tc superconductor ($La_{2-x}Sr_xCuO_4$) has been described in terms of the transfer time $\tau_{transf}$ between two consecutive scattering events along the bond. To account for the presumed polaron formation in the process, the transfer time $\tau_{transf}$ has been introduced in a mode-coupled form. Using a quasi 1-D polaron theory, the relaxation time $\tau_{transf}$ has led to a temperature-dependent dc-conductivity ($\propto 1/\tau_{transf}$) that proved in a quantitative concert with the experimental data on the axial conductivity at various doping levels, and perhaps less so on the in-plane conductivity. For a better agreement with experiment, one-phonon scattering processes have also been invoked. The resulting improvement of the theory has made the fits to the experimental temperature dependences nearly perfect. However, the barriers along the configurational path obtained as fitting parameters of the phonon-coupled transfer have proved insensitive to the doping level which by far is less than convincing. Among other things, it implies that the off-center parameters are not dependent on the site occupancy Further improvements of the theory might be expected on incorporating a quantum statistics in lieu of the classical statistics used so far for dealing with rate processes [32].

Following the above suggestions, we define the dc-conductivity dominated by scattering from double-well potentials:

$$\sigma(0,T) = en_{pol}(a_{bond}/\tau_{transf}) = en_{pol} a_{bond} k_{transf} \qquad (44)$$

where $n_{pol}$ is the polaron density, $a_{bond}$ is the Cu-O bond length and $k_{transf} = 1/\tau_{transf}$ is the polaron tranfer rate along the bond. The double-well character (1-D site-splitting) is all included in the relaxation time $\tau_{transf}$ applying to a particular situation [15]. The polaron density $n_{pol}$ in equation (44) is sensitive to the electronic configuration in that it may incite electronic transitions either between two localized levels (a donor-acceptor transition) or from a localized level to a polaron band. In the latter case the polaron density with Boltzmann-tail statistics is dependent of the effective density of states in the polaron conductive band giving

$$n_{pol} = [g_c \, 2(2\pi \, m_{pol} \, k_B T / h^2)^{3/2}]^{-1} \exp(-\varepsilon_b / 2k_B T) \, x \, \Pi \qquad (45)$$

where $g_c$ is the number of valleys in the polaron band, $m_{pol}$ is the polaron mass, $\varepsilon_b$ is the polaron binding energy, $x$ is the doping level, $\Pi$ is the density of polaron sites.

We now see just how the off-center character is reflected on the optical spectra: it simply modulates the spectral magnitude scale-wise without affecting the shape. As the temperature T is raised above zero, the spectral magnitude is first very slowly varying with T which is then followed by a transition range, where $\sigma(\nu,T)$ is gaining strength, before entering the steep thermally-activated range where the spectral magnitude follows suit [15]. At this point we stress that conclusions on the thermal constancy of an absorption intensity drawn from low-temperature measurements should be extended cautiously. However, if an absorption band starts rising more or less steeply after remaining relatively insensitive to T at lower T, the off-center character of the underlying absorbing centers is a likely possibility. This is eventually the clearest off-site fingerprint to be taken from an optical absorption spectrum.

As stated above, the $\sigma(\nu,T)/\sigma(0,T)$ ratio in equation (43) describes a band peaking at $2\pi\hbar\nu_{peak} = 2\varepsilon_b$. This enables one to proceed as follows: We first determine $\varepsilon_b$ from the experimental peak energy then use it for calculating the frequency-dependent factor $\sigma(\nu,T)/\sigma(0,T)$ by equation (43) at any given temperature T. Having done this, we use the experimental $\sigma(\nu,T)$ temperature data to calculate $\sigma(0,T)$. We finally analyze the temperature dependence of $\sigma(0,T)$ thus obtained to search for off-center fingerprints. Among these are the intersite barrier, the coupled phonon frequency, and the intersite electron coupling energy [15]. The fingerprint-seeking steps are illustrated in Figure 3 (a)-(c) for the axial dc-conductivity in $La_{2-x}Sr_xCuO_4$ and the associated IR optical spectrum.

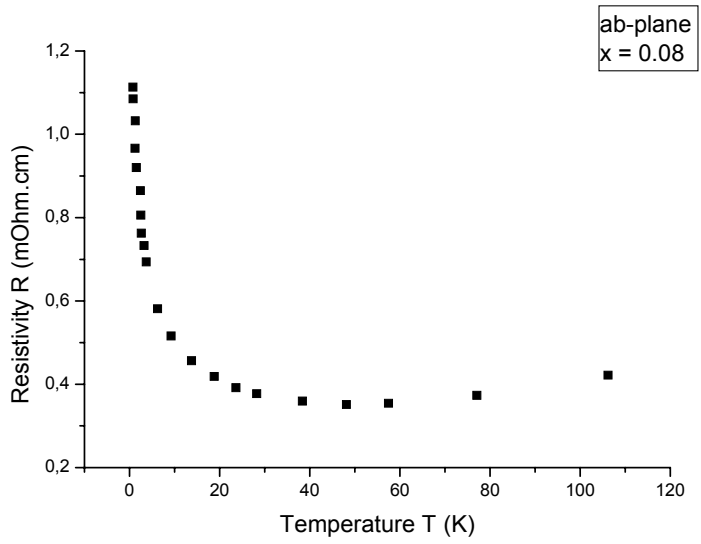

Figure 3 (a)

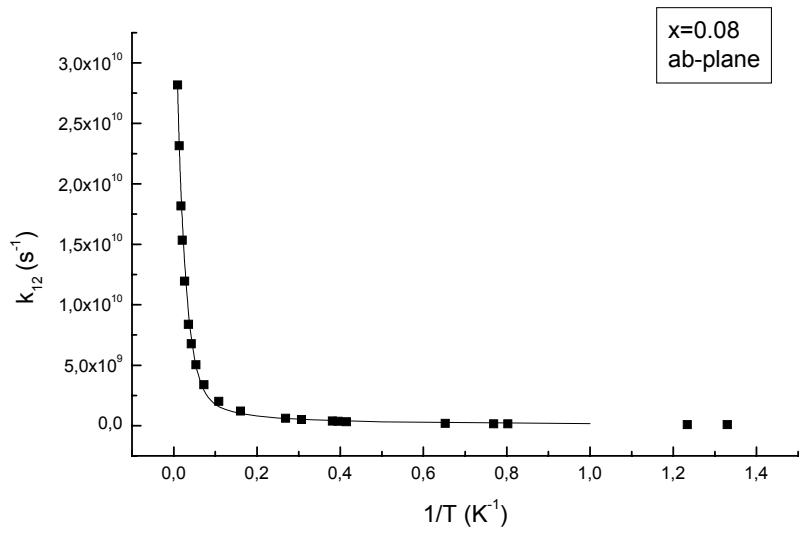

Figure 3 (b)

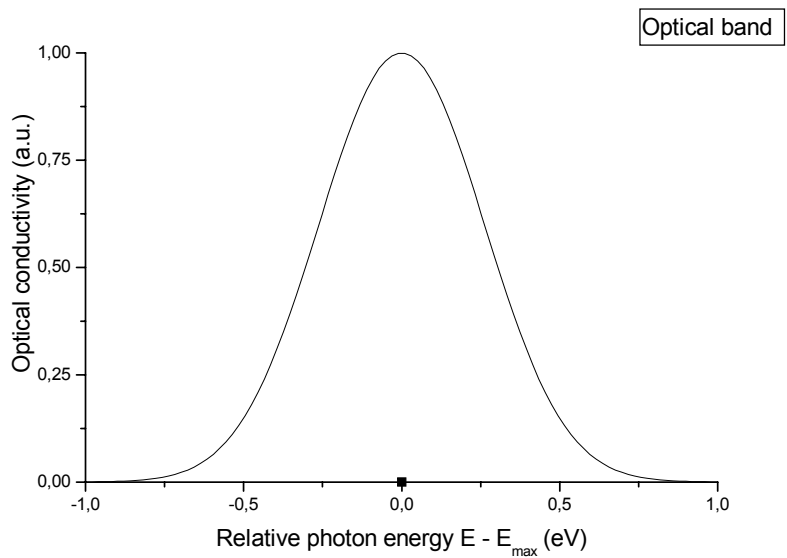

Figure 3 (c)

Figure 3. Illustrating the fingerprint seeking steps for the in-plane optical conductivity of $La_{2-x}Sr_xCuO_4$: (a)- dc-conductivity $\sigma(0,T)$ (experimental); (b)- temperature dependence of the $O(A_1) \leftrightarrow Cu(P) \leftrightarrow O(A_2)$ rate constant, as derived from the conductivity data in (a); (c)- IR optical band $\sigma(\nu,T)/\sigma(0,T)$ calculated using equation (43) at $E_{max} \sim 1$ eV. The function in (b) is typical for the vibronic polaron transfer scattered at a double-well.

The relaxation rates of the rotating off-center $Li^+$ impurity in KCl has been measured in addition to the FIR optical band [10] and will be dealt with along similar lines later.